\documentclass[aps,pra,twocolumn,superscriptaddress]{revtex4-1}

\usepackage{amsfonts,amssymb,amsmath}
\usepackage{mathtools}
\usepackage[]{graphics,graphicx,epsfig}
\usepackage{amsthm,multirow}
\begin{document}

\title{Quantum phase estimation using a multi-headed cat state}

\author{Su-Yong \surname{Lee}}
\affiliation{Centre for Quantum Technologies, National University of Singapore, 3 Science Drive 2, 117543 Singapore, Singapore}

\author{Chang-Woo Lee}
\affiliation{Department of Physics, Texas A\&M University at Qatar, Education City, POBox 23874, Doha, Qatar}
\affiliation{School of Computational Sciences, Korea Institute for Advanced Study, Hoegi-ro 85,Dongdaemun-gu, Seoul 130-722, Korea}

\author{Hyunchul Nha}
\affiliation{Department of Physics, Texas A\&M University at Qatar, Education City, POBox 23874, Doha, Qatar}

\author{Dagomir \surname{Kaszlikowski}}
\affiliation{Centre for Quantum Technologies, National University of Singapore, 3 Science Drive 2, 117543 Singapore, Singapore}
\affiliation{Department of Physics, National University of Singapore, 2 Science Drive 3, 117542 Singapore, Singapore}

\date{\today}

\begin{abstract}
It was recently shown that an entangled coherent state, which is a superposition of two different coherent states, can surpass the performance of noon state in estimating an unknown phase-shift. This may hint at further enhancement in phase estimation by incorporating more component states in the superposition of resource state. We here introduce a four-headed cat state (4HCS), a superposition of four different coherent states, and propose its application to quantum phase estimation. We demonstrate the enhanced performance in phase estimation by employing an entangled state via the 4HCS, which can surpass that of the two-headed cat state (2HCS), particularly in the regime of small average photon numbers.
Moreover, we show that an entangled state modified from the 4HCS can further enhance the phase estimation, even in the regime of large average photon number under a photon-loss channel. Our investigation further extends to incorporate an increasingly large number of component states in the resource superposition state and clearly show its merit in phase estimation. 
\end{abstract}

%\pacs{42.50.St, 42.50.Dv, 06.20.Dk}
\maketitle

\section{Introduction}

In quantum mechanics, the superposition principle provides a crucial basis to observe phenomena beyond the predictions of classical physics. One example is a quantum entangled state that can exhibit stronger correlation than classically possible. Among many different superposition states, the superposition of coherent states (SCSs) with the same amplitude but two different phases has been a subject of great interest for decades. Its constituent states can be macroscopically distinguishable in the limit of large amplitude, and the SCS may become an important tool to study a lot of fundamental issues, e.g. the decoherence of macroscopic superposition state. Numerous efforts have been devoted to identifying feasible schemes to generate SCSs,
together with experimental achievements in, e.g., a trapped $^9$Be$^+$ ion system \cite{MMKW96} and a Bose-Einstein condensate (BEC) with Rb atoms \cite{GMHB02}. 
For the latter case, the dynamics of quantum decoherence was also observed in a high \textit{Q} microwave cavity \cite{B96}
and theoretical proposals were made to generate a macroscopic superposition of phase states in BEC \cite{PPS08} and in Bose Josephson junction \cite{FMH09}.

%Review of entangled coherent states \cite{S12}
An entangled coherent state (ECS) is an entangled version of SCS, of which a typical form can be expressed by 
$|\mathrm{ECS}\rangle=\mathcal{N}(|\alpha\rangle|-\alpha\rangle \pm |-\alpha\rangle|\alpha\rangle)$,
%\begin{equation}
%\ket{\mathrm{ECS}} = \mathcal{N} [\ket{\alpha}\ket{\alpha} \pm \ket{-\alpha}\ket{-\alpha}]
%\end{equation}
with $ \mathcal{N} $ a normalization factor. ECSs can readily be produced by feeding single-mode SCSs into a beam splitter, or by employing nonlinear Kerr medium \cite{Mecozzi87,Yurke87,Sanders92,Gerry99} and atomic ensemble in a cavity \cite{Paternostro03}. There are also probabilistic heralding schemes using photon subtraction or homodyne detections \cite{Ourjoumtsev09,Brask10}.
ECSs have a wide scope of applications ranging from fundamental tests of quantum mechanics \cite{Mann95,Paternostro10,McKeown11,Lee13}
to practical quantum computation \cite{Cochrane99,JKL01,JK02,Ralph03,Lund08,Kim10,Lund11}, communication \cite{Sangouard10,Neergaard13,Ralph09}, and metrology \cite{Munro02,Joo11}.
%In particular, ECSs are essential resource for  SCS-based quantum computation and communication scheme \cite{Ralph09}.

There have also been some studies to extend the quantum superposition to involve more than two component coherent states.
The SCS with three different phases was considered to produce superposition of three photons \cite{Y13}. Furthermore, the SCS with four different phases was proposed in a microwave cavity quantum electrodynamics (CQED) of bang-bang quantum Zeno dynamics (QZD) control \cite{R10}, using initial even coherent states. Recently the SCS with up to four different phases was implemented in a CQED architecture, 
using off-resonant interactions between a waveguide cavity resonator and a superconducting transmon qubit \cite{V13}. On an application side, encoding a logical qubit in a SCS with four different phases was proposed to protect the logical qubit against relaxation \cite{L13}.
Moreover, SCSs with different amplitudes was proposed to generate mesoscopic field state superpositions,
e.g., $(|4\rangle+|4i\rangle+|3e^{i5\pi/4}\rangle+|0\rangle)/2$, where the complex values in the kets represent different amplitudes of coherent states \cite{R12}.

Recently, Joo {\it et al.} found that an ECS with two component coherent states can provide a better sensitivity in estimating an unknown phase shift in the Mach-Zehnder interferometer than the so-called noon state \cite{Joo11}. In order to achieve further enhancement in phase sensitivity, it seems worthwhile to extend approach to superimposing more component states in the resource state.
In this paper we study the SCS with four different phases, which is referred to as a four-headed cat state (4HCS).
The 4HCS manifests more interference in its phase space distribution than an even coherent state, and it is thus interesting to investigate its nonclassical properties in detail, partcularly using the nonclassical measures including sub-Poissonian statistics, negativity in phase space, and degree of entanglement potential. Specifically, we show that entangled resources employing 4HCSs can enhance resolution in phase estimation in the regime of small average photon number. Furthermore we propose a modified entangled state that can provide the enhancement of the phase estimation even in the regime of large average photon number under a photon-loss channel. Our investigation further extends to include an increasingly large number of component states in the superposition of resurce state and demonstrate its advantage in phase estimation. 

This paper is structured as follows. In Sec. II, we introduce the SCS with four different phases. In Sec. III, we propose to use an entangled state produced by injecting the 4HCS into a beam splitter for phase estimation in Mach-Zehnder interferometer and demonstrate its advantage over the entangled state by the 2HCS.  Furthermore, we introduce a modified entangled state and multi-component superposition states to observe the enhancement of phase sensitivity. We summarize our main results in Sec. IV.

\section{Four-headed cat state}
The SCS comprising $N$ coherent states of the same amplitude $\alpha$ but evenly-distributed phases $2\pi\frac{n}{N}$ ($n=0,1,\ldots,N-1$) can be represented by
%\begin{eqnarray}
%|SC\rangle &\equiv&\frac{1}{\sqrt{M}}\sum^{N-1}_{n=0}|\alpha e^{i2\pi\frac{n}{N}}\rangle \nonumber\\
%&=&\frac{N}{\sqrt{M}}e^{-\frac{|\alpha|^2}{2}}\sum^{\infty}_{n=0}\frac{\alpha^{N\cdot n}}{\sqrt{(N\cdot n)!}} |N\cdot n\rangle,
%\end{eqnarray}
\begin{eqnarray}
|C_N(\alpha)\rangle &\equiv&\frac{1}{\sqrt{\mathcal{M}_N(\alpha)}}\sum^{N-1}_{n=0}|\alpha e^{i2\pi\frac{n}{N}}\rangle \nonumber\\
&=&\frac{Ne^{-\frac{|\alpha|^2}{2}}}{\sqrt{\mathcal{M}_N(\alpha)}}\sum^{\infty}_{n=0}\frac{\alpha^{N\cdot n}}{\sqrt{(N\cdot n)!}} |N\cdot n\rangle,
\end{eqnarray}
where $\mathcal{M}_N(\alpha)$ is the normalization constant. The component state $|N\cdot n\rangle$ in the second line of Eq. (1) is a Fock state of number $Nn$. For each $N=2,3,\ldots$, the SCS
is an eigenstate of the $N$th-order annihilation operator, i.e., %$\hat{a}^N|SC\rangle=\alpha^N|SC\rangle$.
$\hat{a}^N|C_N(\alpha)\rangle=\alpha^N|C_N(\alpha)\rangle$.
For $N=2$, the SCS is an even coherent state $(|\alpha\rangle+|-\alpha\rangle)$, which is also called {\it two-headed cat state} (2HCS).
Here we are interested in a superposition of coherent states with four different phases ($N=4$) referred to as {\it four-headed cat state} (4HCS). This corresponds to
\begin{eqnarray}
|C_4(\alpha)\rangle &=&\frac{1}{\sqrt{\mathcal{M}_4}}(|\alpha\rangle +|\alpha e^{i\frac{\pi}{2}}\rangle+|\alpha e^{i\pi}\rangle+|\alpha e^{i\frac{3\pi}{2}}\rangle)\nonumber\\
&=&\frac{4e^{-\frac{|\alpha|^2}{2}}}{\sqrt{\mathcal{M}_4}}\sum^{\infty}_{n=0}\frac{\alpha^{4\cdot n}}{\sqrt{(4\cdot n)!}} |4\cdot n\rangle,
\end{eqnarray}
with %$M_4=4[1+e^{-2|\alpha|^2}+2e^{-|\alpha|^2}\cos(|\alpha|^2)]$ 
$\mathcal{M}_4= 4[1+e^{-2|\alpha|^2}+2e^{-|\alpha|^2}\cos(|\alpha|^2)]$. 
Due to more pronounced interference structure in its phase-space distribution, the 4HCS may show stronger nonclassical effects than the 2HCS.
In optimal phase estimation, thus, the entangled state that is generated by using 4HCS may be more useful than the entangled state by a 2HCS \cite{LJL13}.

\section{Optimal phase estimation}
We study an optical phase measurement and investigate its sensitivity to beat the shot-noise limit \cite{C81,D08} by using a 4HCS.
%{\bf We expect that the 4HCS provides even better performance than a 2HCS in the optimal phase estimation, due to the more pronounced multiple interference in its phase space.} {\bf Note by Nha: I suggest to remove this bold-faced paragraph. Physically, there is no logical link between the sensitivity in phase estimation and the interference structure in phase-space distribution. Do we have justification? Otherwise, it suffices to remove it with no harm.} 
The sensitivity in a phase measurement can be obtained by considering a Mach-Zehnder interferometer, which is designed to estimate the phase difference $\phi$ between two optical paths.
We here show that the measurement sensitivity is enhanced by employing the 4HCS as an input to the interferometer.

The problem of phase estimation may be addressed in terms of phase uncertainty $\delta\phi_c\geq 1/\sqrt{F_Q}$ for a single-shot measurement \cite{ZPK10}, where $F_Q$ is the quantum Fisher information.
It was recently shown that the optimal phase estimation employing a 2HCS performs better than that employing a noon state 
under the same energy constraint \cite{Joo11,J12},
where a phase shift operation was applied to one of the optical paths. 
On the other hand, it has also been discovered that the Fisher information may give unequal results depending on whether the phase shift is made on only one mode or on two modes, even though the phase differences in those two configurations are the same \cite{JD12}. This may leave some controversy, but Jarzyna and Demkowicz-Dobrza\'nski pointed out that the problem can be attributed to the lack of external phase reference \cite{JD12}. Therefore, instead of considering a pure resource state $\rho_{in}$, they suggested to use a phase-averaged state $\rho_m$ to obtain a reasonable Fisher information, i.e., 
\begin{eqnarray}
\rho_m=\int^{2\pi}_0 \frac{d\theta}{2\pi}V^a_{\theta}V^b_{\theta}\rho_{in}V^{a \dag}_{\theta}V^{b \dag}_{\theta},
\end{eqnarray}
where $V^x_{\theta}=\exp(-i\theta \hat{x}^{\dag}\hat{x})$. In this case, the Fisher information only depends on the phase difference regardless of experimental configurations.

In this section, we calculate the Fisher information using both a pure resource state and a phase-averaged state, with phase-shifts made on one-mode and two-modes, respectively, for comparison purpose.
We first apply a phase shift operation to one of the optical paths ($e^{i\phi\hat{b}^{\dag}\hat{b}}$). We then compare the 4HCS with the 2HCS as a resource to phase measurement and demonstrate the advantage of using 4HCS.
We also propose a modified entangled state from the one by 4HCS, which can further enhance the phase sensitivity.
%Then, we also consider a phase averaged state of the modified entangled state, which is shown to enhance the phase sensitivity even in the presence of photon loss.

\subsection{Four-headed cat state}

\begin{figure}
\vspace{-1.1in}
\centerline{\scalebox{0.4}{\includegraphics[angle=0]{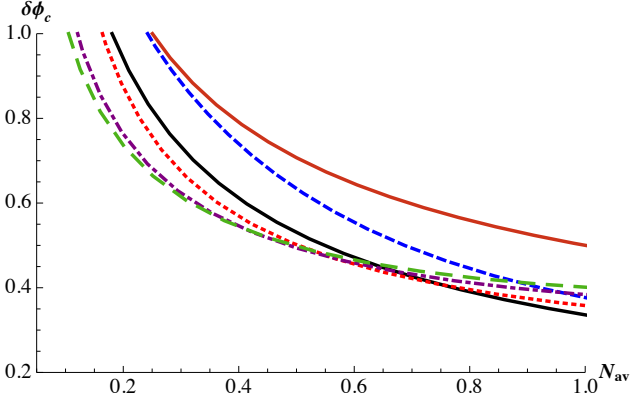}}}
\vspace{-1.1in}
\caption{Optimal phase estimation as a function of the average photon number for mode $a$, that is, $N_{av}$, using entangled states produced by injecting the 4HCS and an additional coherent state ($|\frac{\beta}{\sqrt{2}}\rangle$) for $\beta=\alpha$ (blue dashed), $\alpha/2$ (red dotted), $\alpha/4$ (purple dot-dashed), 
$0$ (green long-dashed), and an entangled coherent state (black solid).
The classical limit is given by a coherent state (red solid).
}
\label{fig:fig1}
\end{figure}
Here we consider the 4HCS and the 2HCS under the same average photon number as a resource to generate an entangled state. 
The entangled state is produced by injecting the 4HCS (2HCS) and an additional coherent state into a 50:50 beam splitter. The resulting state is used as an input to the Mach-Zehnder interferometer to estimate the phase difference $\phi$ between two optical paths, the quality of which can be quantified by the quantum Fisher information.

Injecting the 4HCS %($|\frac{4cat}{\sqrt{2}}\rangle$) 
$|C_4 (\alpha/\sqrt{2})\rangle$
and a coherent state %($|\frac{\beta}{\sqrt{2}}\rangle$)
$|\beta/\sqrt{2}\rangle$
into a 50:50 beam splitter, we obtain an entangled state
\begin{eqnarray}
|\psi_{in}\rangle&=&\frac{1}{\sqrt{M_e}}(|B_{+}\rangle_a|B_{-}\rangle_b+|B_{-}\rangle_a|B_{+}\rangle_b\nonumber\\
&&+|D_{+}\rangle_a|D_{-}\rangle_b+|D_{-}\rangle_a|D_{+}\rangle_b),
\end{eqnarray}
where $B_{\pm}=(\beta\pm\alpha)/2$, $D_{\pm}=(\beta\pm i\alpha)/2$, and %$M_e=4(1+e^{-|\alpha|^2}+2e^{-\frac{|\alpha|^2}{2}}\cos(\frac{|\alpha|^2}{2}))$.
$M_e= \mathcal{M}_4 (\alpha/\sqrt{2})$.
%\subsubsection{Single phase shift}
Then, applying a phase shift operation to one of the optical paths $|\psi_{out}\rangle=(I\otimes e^{i\phi\hat{b}^{\dag}\hat{b}})|\psi_{in}\rangle$, the quantum Fisher information is given by
\begin{eqnarray}
F_{Q,1}&=&4(\langle \psi^{'}_{out}|\psi^{'}_{out}\rangle-|\langle \psi^{'}_{out}| \psi_{out}\rangle|^2),\nonumber\\
&=& 4(\langle \psi_{in}|\hat{n}^2_b|\psi_{in}\rangle-\langle \psi_{in}|\hat{n}_b|\psi_{in}\rangle^2),
\end{eqnarray}
where $|\psi^{'}_{out}\rangle=\partial |\psi_{out}\rangle/\partial \phi$ and $\hat{n}_b=\hat{b}^{\dag}\hat{b}$. 
The lower bound of phase uncertainty, $1/\sqrt{F_{Q,1}}$, is then given by the number variance of one of the two input modes \cite{Bowen}.
We obtain
\begin{eqnarray}
\langle \psi_{in}|\hat{n}^2_b|\psi_{in}\rangle&=&\frac{1}{M_e}[(\frac{|\alpha|^4+|\beta|^4}{4}+|\beta|^2)(1+e^{-|\alpha|^2})\nonumber\\
&&+|\alpha|^2(1+|\beta|^2)(1-e^{-|\alpha|^2})\nonumber\\
&&+2e^{-\frac{|\alpha|^2}{2}} \{ (|\beta|^2+\frac{|\beta|^4-|\alpha|^4}{4})\cos{\frac{|\alpha|^2}{2}}\nonumber\\
&&-|\alpha|^2(1+|\beta|^2)\sin{\frac{|\alpha|^2}{2}}\}],\\
%\end{eqnarray}
%\begin{eqnarray}
\langle \psi_{in}|\hat{n}_b|\psi_{in}\rangle &=&\frac{1}{M_e}[|\alpha|^2(1-e^{-|\alpha|^2}-2e^{-\frac{|\alpha|^2}{2}}\sin{\frac{|\alpha|^2}{2}})\nonumber\\
&&+|\beta|^2(1+e^{-|\alpha|^2}+2e^{-\frac{|\alpha|^2}{2}}\cos{\frac{|\alpha|^2}{2}})]. \nonumber
\end{eqnarray}
The average photon number for the input mode $a$ is given by $N_{av}=\langle \psi_{in}|\hat{n}_a|\psi_{in}\rangle=\langle \psi_{in}|\hat{n}_b|\psi_{in}\rangle$,
where $\hat{n}_a=\hat{a}^{\dag}\hat{a}$. The total photon number of two modes is twice as large as $N_{av}$ for a symmetric state, like the ones considered throughout this paper. We thus use $N_{av}$ as the energy constraint from now on.

For the 2HCS, we consider only the case in which the additional coherent state is fixed as %$|\frac{\alpha}{\sqrt{2}}\rangle$.
$|\alpha/\sqrt{2}\rangle$.
Then, an entangled state is produced by injecting the 2HCS %($|\frac{\alpha}{\sqrt{2}}\rangle$)
$|C_2 (\alpha/\sqrt{2})\rangle$
and a coherent state %($|\frac{\alpha}{\sqrt{2}}\rangle$) 
$|\alpha/\sqrt{2}\rangle$
into a 50:50 beam splitter, which yields an entangled coherent state ($|\alpha\rangle_a|0\rangle_b+|0\rangle_a|\alpha\rangle_b$) previously studied in \cite{Joo11,J12}.
After applying a phase shift operation to one of the optical paths, 
we obtain the quantum Fisher information \cite{Joo11} 
\begin{eqnarray}
F_{Q,1}=\frac{2(|\alpha|^2+|\alpha|^4)}{1+e^{-|\alpha|^2}}-\frac{|\alpha|^4}{(1+e^{-|\alpha|^2})^2}.
\end{eqnarray}
The average photon number for mode $a$ is given by $N_{av}=|\alpha|^2/2(1+e^{-|\alpha|^2})$.

We examine how the coherent state $|\beta/\sqrt{2}\rangle$ injected as the other input into the beam-splitter 
can be adjusted to the optimal phase estimation by the 4HCS, 
in order to beat the optimal phase estimation by the 2HCS. In Fig. 1, we plot the  phase sensitivity as a function of the input energy $N_{av}$, for the cases of the 4HCS and the 2HCS.
Note that the classical limit of the optimal phase estimation is obtained by an input coherent state
$|\psi_{in}\rangle=|\alpha/\sqrt{2}\rangle_a|\alpha/\sqrt{2}\rangle_b$, as shown in Fig. 1.
In the regime of $\beta \leq \alpha/2$, we see that the 4HCS gives the enhancement of the optimal phase estimation over the 2HCS within the small average photon number $N_{av}\leq 0.7$. It is explained that, in the regime of $N_{av}\leq 0.7$, the variance of mode b in Eq. (4) is larger than that in the entangled coherent state ($|\alpha\rangle_a|0\rangle_b+|0\rangle_a|\alpha\rangle_b$) under the same average photon number.
The enhancement regime of the average photon number is rather small. 
Instead of injecting the 4HCS into a 50:50 beam splitter, thus, we can consider its modified version, i.e., 
the superposition of two different coherent states in one mode  or another.
Note that, in the case of the 2HCS, one coherent state is in one mode or another.
We next investigate this input entangled state to enhance the phase sensitivity in a larger regime of $N_{av}$.

\subsection{Modified entangled state}
\begin{figure}[htbp]
\vspace{0.0in}
\centering{\scalebox{0.45}{\includegraphics[angle=0]{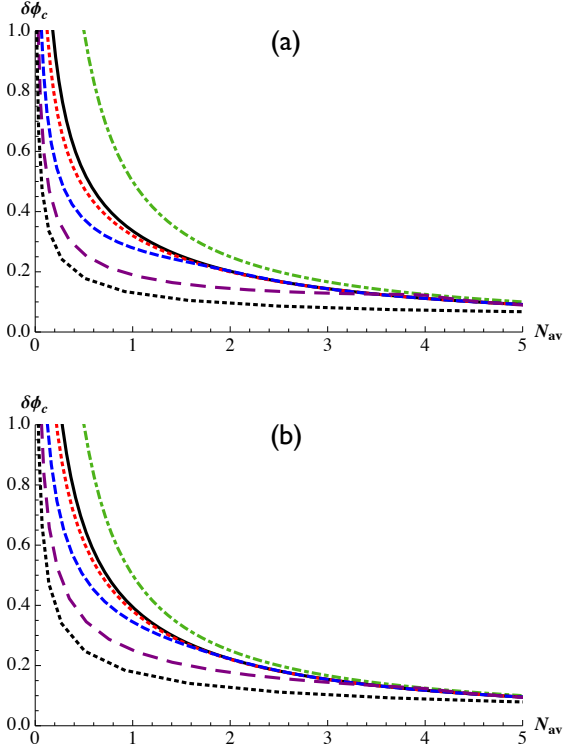}}}
\vspace{-0.5in}
\caption{Phase sensitivity as a function of the average photon number $N_{av}$ using the noon state (green dot-dashed), the entangled coherent state 
($|\alpha\rangle_a|0\rangle_b+|0\rangle_a|\alpha\rangle_b$) (black solid), the modified entangled state (Eq. (8))
(red dotted), and the extended entangled state $(|C_N(\alpha)\rangle_a|0\rangle_b+|0\rangle_a|C_N(\alpha)\rangle_b)$ for 
$N=4$ (blue dashed), $8$ (purple long-dashed), and $16$ (black dotted).
(a) Pure states, (b) Phase-averaged states of the pure states. 
}
\label{fig:fig2}
\end{figure}

We propose a modified entangled state to enhance the optimal phase estimation in the regime of $N_{av}>1$.
Similar to a noon state, the modified entangled state may be proposed as
\begin{eqnarray}
|\psi_{in}\rangle=\frac{(|\alpha\rangle_a+|-\alpha\rangle_a)|0\rangle_b+|0\rangle_a(|\alpha\rangle_b+|-\alpha\rangle_b)}
{2(1+e^{-|\alpha|^2})},
\end{eqnarray}
where the superposition of two different coherent states is in one mode or another.
After applying the phase shift operation to the input state $|\psi_{out}\rangle=(I\otimes e^{i\phi\hat{b}^{\dag}\hat{b}})|\psi_{in}\rangle$, 
we evaluate its quantum Fisher information with the quantities
\begin{eqnarray}
\langle \psi_{in}|\hat{n}^2_b|\psi_{in}\rangle&=&\frac{|\alpha|^2[1+|\alpha |^2+(|\alpha |^2-1)e^{-2|\alpha|^2}]}{2(1+e^{-|\alpha|^2})^2}
,\nonumber\\
\langle \psi_{in}|\hat{n}_b|\psi_{in}\rangle &=& \frac{|\alpha|^2(1-e^{-2|\alpha|^2})}{2(1+e^{-|\alpha|^2})^2}.
\end{eqnarray}
In Fig. 2 (a), we plot the optimal phase estimation as a function of $N_{av}$ for the modified entangled state in Eq. (8).  
%Note that, at $N=1$, the modified entangled state is equal to the entangled state by the 2HCS.
For comparison, the equivalent amplitude and $N_{av}$ of the noon state are given by choosing $n=|\alpha|^2$ and $N_{av}=n/2$, respectively \cite{Joo11}.
We find that the modified entangled state provides the enhancement of the optimal phase estimation over the entangled coherent state, even in the regime of $N_{av}>1$. Furthermore, extending the superposition to $N$ different coherent states in Eq. (8), i.e., $|C_N(\alpha)\rangle_a|0\rangle_b+|0\rangle_a|C_N(\alpha)\rangle_b$, the optimal phase estimation is further enhanced with the number $N$, as shown in Fig. 2 (a).
For the superposition of $N$ different coherent states, the detail calculations are given in Appendix.
At $N=1$ , the extended entangled state is equivalent to the entangled coherent state. At $N=2$, it is the same as Eq. (8).
It is explained that, under the same average photon number, the variance of $|C_N(\alpha)\rangle$ increases with the number $N$. Quantitatively, the ratio $F_{Q,1}/N_{av}$ is described with $4(1+Q_{|C_N(\alpha)\rangle})$, where $Q_{|C_N(\alpha)\rangle}$ is the Mandel Q-factor of $|C_N(\alpha)\rangle$. Thus, the Mandel Q-factor increases with the number $N$ even if it is not always true for the whole regime of $|\alpha|$.
It is expected that the phase sensitivity is improved even more with higher $N$.

\subsubsection{Generation scheme}
In Fig. 3, the modified entangled state of Eq. (8) can be produced by injecting two two-headed cat states into a 50:50 beam splitter, 
\begin{eqnarray}
&&(|\frac{\alpha}{\sqrt{2}}\rangle_a+|\frac{-\alpha}{\sqrt{2}}\rangle_a)
(|\frac{\alpha}{\sqrt{2}}\rangle_b+|\frac{-\alpha}{\sqrt{2}}\rangle_b) \nonumber\\
&&\xrightarrow{50:50 BS} 
(|\alpha\rangle_a+|-\alpha\rangle_a)|0\rangle_b+|0\rangle_a(|\alpha\rangle_b+|-\alpha \rangle_b).\nonumber
\end{eqnarray}
Then, applying a series of conditional phase shift (CPS) operations on each output mode, we obtain the extend entangled state, 
$|C_N(\alpha)\rangle_a|0\rangle_b+|0\rangle_a|C_N(\alpha)\rangle_b$. 
The CPS operation was implemented in a superconducting transmon qubit coupled to a waveguide cavity resonator \cite{V13}.
The CPS operation is implemented by $\hat{C}\equiv I\otimes|H\rangle\langle H|+e^{i\varphi \hat{a}^{\dag}\hat{a}}\otimes|V\rangle\langle V|$. 
Applying the CPS operation on target and ancillary qubits, we represent the output qubits as
\begin{eqnarray}
&&\hat{C}|\alpha\rangle\otimes (|H\rangle+|V\rangle)=|\alpha\rangle|H\rangle+|\alpha e^{i\varphi}\rangle|V\rangle\nonumber\\
&&\xrightarrow{|+\rangle} |\alpha\rangle+|\alpha e^{i\varphi}\rangle,
\end{eqnarray}
where $|\pm\rangle=(|H\rangle\pm|V\rangle)/\sqrt{2}$. Detecting the ancillary qubits as $|+\rangle$ on each output mode, we obtain an extended entangled state. Repeating the process with different phase component ($\varphi$), we produce the extended entangled state $(|C_N(\alpha)\rangle_a|0\rangle_b+|0\rangle_a|C_N(\alpha)\rangle_b)$ at $N=2^{k+1}$ ($k$: iteration time), as shown in Fig. 3.
The phase component ($\varphi$) changes into $\varphi_k=2\pi/2^{k+1}$ with $k$ iterations.

\begin{figure}
\centerline{\scalebox{0.37}{\includegraphics[angle=0]{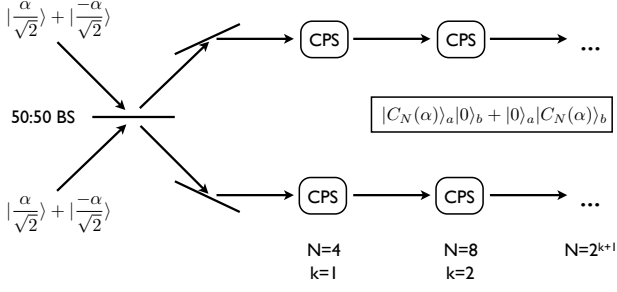}}}
\vspace{-1.6in}
\caption{Generation scheme of the modified and extended entangled states. $k$ is the iteration time for the conditional phase shift (CPS) operation.
}
\label{fig:fig3}
\end{figure}

\subsection{Phase-averaged state}
If we apply a phase shift operation on the two optical paths $|\psi_{out}\rangle=(e^{-i\frac{\phi}{2}\hat{a}^{\dag}\hat{a}} \otimes e^{i\frac{\phi}{2}\hat{b}^{\dag}\hat{b}})|\psi_{in}\rangle$, the quantum Fisher information is given by
\begin{eqnarray}
F_{Q,2}=\langle \psi_{in}|(\hat{n}_b-\hat{n}_a)^2|\psi_{in}\rangle-\langle \psi_{in}|(\hat{n}_b-\hat{n}_a)|\psi_{in}\rangle^2. \nonumber\\
\end{eqnarray}
The optimal phase uncertainty $1/\sqrt{F_{Q,2}}$ is now given by the variance of number difference between the two input modes. In general, the quantum Fisher information for a pure input state may depend on the configuration of phase operations, thus giving $F_{Q,2}\neq F_{Q,1}$, even though the phase difference is the same \cite{JD12}. This inconsistency can be resolved if one considers a phase averaged state as an input, i.e., Eq. (3).
%\begin{eqnarray}
%\rho_m=\int^{2\pi}_0 \frac{d\theta}{2\pi}V^a_{\theta}V^b_{\theta}\rho_{in}V^{a \dag}_{\theta}V^{b \dag}_{\theta},
%\end{eqnarray}
%where $V^x_{\theta}=\exp(-i\theta \hat{x}^{\dag}\hat{x})$ \cite{JD12}. 
Furthermore, the phase-averaged state makes a practical sense when one has no access to external phase reference. After applying a phase shift operation to the phase averaged state, 
we obtain the quantum Fisher information.
For a mixed state, the quantum Fisher information can be obtained by the diagonalization of the mixed state \cite{KSD11,ZLYJ13,LJW13} as 
\begin{eqnarray}
F_{q}=4\sum_i \lambda_i f_{i}
-\sum_{i\neq j}\frac{8\lambda_i \lambda_j}{\lambda_i+\lambda_j}|\langle \lambda^{'}_i|\lambda_j\rangle|^2, 
\end{eqnarray}
where $\lambda_i$ and $ |\lambda_i\rangle$ are the eigenvalues and eigenvectors of the mixed state $\rho_{ab}=\sum_i\lambda_i|\lambda_i\rangle\langle\lambda_i|$,
$f_{i}=(\langle \lambda^{'}_i|\lambda^{'}_i\rangle-|\langle \lambda^{'}_i|\lambda_i\rangle|^2)$,
and $|\lambda^{'}_i\rangle=\partial |\lambda_i\rangle/\partial \phi$.

We obtain a phase averaged state of the modified entangled state as
\begin{eqnarray}
\rho_m=\frac{e^{-|\alpha|^2}}{(1+e^{-|\alpha|^2})^2}\sum^{\infty}_{n=0}\frac{|\alpha|^{2n}}{n!}[1+(-1)^n] 
|noon\rangle_{ab}\langle noon|,\nonumber\\
\end{eqnarray}
where $|noon\rangle_{ab}=(|n\rangle_a|0\rangle_b+|0\rangle_a|n\rangle_b)/\sqrt{2}$. On applying a phase shift operation, 
we obtain the quantum Fisher information
\begin{eqnarray}
F_q=\frac{|\alpha|^2}{(1+e^{-|\alpha|^2})^2}[ 1+|\alpha|^2+(|\alpha|^2-1)e^{-2|\alpha|^2}].
\end{eqnarray}

On the other hand, the entangled coherent state after phase-averaging is given by
\begin{eqnarray}
\rho_m=\frac{e^{-|\alpha|^2}}{1+e^{-|\alpha|^2}}\sum^{\infty}_{n=0}\frac{|\alpha|^{2n}}{n!}|noon\rangle_{ab}\langle noon|,
\end{eqnarray}
whose quantum Fisher information becomes
\begin{eqnarray}
F_q=\frac{|\alpha|^2(1+|\alpha|^2)}{1+e^{-|\alpha|^2}}.
\end{eqnarray}
For a noon state, there is no relative phase between the two modes so that the quantum Fisher information is unchanged under phase-averaging, $F_q=n^2$.
We note that all the above phase-averaged states incidentally give $F_q=F_{Q,2}$, the same as $F_{Q,2}$ in Eq. (11), which is obtained using a pure input state without phase averaging on applying phase shifts on two modes. (Cf. Ref.\cite{JD12})

In Fig. 2 (b), we plot the optimal phase estimation as a function of $N_{av}$, for the phase averaged states in Eqs. (13) and (15).
We see that the enhancement of the optimal phase estimation is similar to the behavior shown in Fig. 2 (a). The modified entangled state can provide the enhancement of the optimal phase estimation over the entangled coherent state, in the regime of $N_{av}>1$.
Extending the superposition to $N$ different coherent states,  
the optimal phase estimation is further enhanced with the number of superposition of coherent states ($N$), as shown in Fig. 2 (b), and it will be improved even more with higher $N$.
The corresponding quantum Fisher information is given in Appendix.

\subsection{Optical loss}
As a final consideration, we show that the phase averaged state of the modified entangled state is useful to enhance the phase estimation even in the presence of optical loss. For simplicity, we assume that each mode experiences the same degree of vacuum noise, which can be modeled by a beam splitter acting on each mode with the same transmission coefficient \cite{Joo11}.

\begin{figure}
\vspace{0.0in}
\centerline{\scalebox{0.45}{\includegraphics[angle=0]{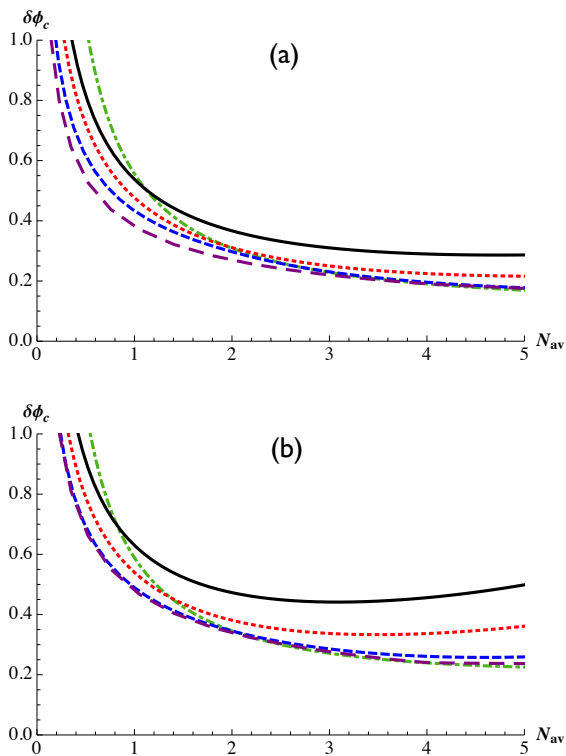}}}
\vspace{-0.5in}
\caption{Phase sensitivity under a loss channel as a function of the average photon number $N_{av}$ using the phase averaged states for the noon state (green dot-dashed), the entangled coherent state $(|\alpha\rangle_a|0\rangle_b+|0\rangle_a|\alpha\rangle_b)$ (black solid), the modified entangled state (Eq. (8)) (red dotted), 
and the extended entangled state $(|C_N(\alpha)\rangle_a|0\rangle_b+|0\rangle_a|C_N(\alpha)\rangle_b)$
for $N=4$ (blue dashed), $8$ (purple long-dashed). (a) T=0.9, (b) T=0.85
($T$: transmission rate of the beam splitters). }
\label{fig:fig4}
\end{figure}

We perform a phase shift operation on one of optical paths in the phase averaged state of the modified entangled state, 
$\rho^{ab}_{out}=(I\otimes e^{i\phi\hat{b}^{\dag}\hat{b}})\rho_m(I\otimes e^{-i\phi\hat{b}^{\dag}\hat{b}})$. 
Subsequently applying a beam splitting operation on each mode, 
$\rho\equiv\hat{B}_{ac}\hat{B}_{bd}\rho_{out}^{ab}\otimes|00\rangle_{cd}\langle 00|\hat{B}^{\dag}_{ac}\hat{B}^{\dag}_{bd}$, and tracing over the additional modes of the state ($\rho_{ab}={\rm Tr}_{cd}[\rho]$), we obtain the output state in terms of its eigenvalues and eigenvectors as
\begin{eqnarray}
\rho_{ab}=\sum^{\infty}_{n=0}(\lambda^+_n|\lambda^+_n\rangle\langle \lambda^+_n|
+\lambda^-_n|\lambda^-_n\rangle\langle \lambda^-_n|),
\end{eqnarray}
where 
\begin{eqnarray}
&&\lambda^{\pm}_n=\frac{e^{-|\alpha|^2}}{2(1+e^{-|\alpha|^2})^2}\frac{(|\alpha|^2T)^n}{n!}[K\pm1\pm(-1)^n], \nonumber\\
&& K=e^{R|\alpha|^2}+(-1)^ne^{-R|\alpha|^2},\nonumber\\
&&|\lambda^{\pm}_n\rangle=\frac{1}{\sqrt{2}}(|n\rangle_a|0\rangle_b\pm e^{in\phi}|0\rangle_a|n\rangle_b),
\end{eqnarray}
with $R=1-T$ ($T$: transmission rate of the beam splitter).

For the phase averaged state of the entangled coherent state, under the same procedures, the eigenvalues and eigenvectors of the output state are given by
\begin{eqnarray}
&&\lambda^{\pm}_n=\frac{(|\alpha|^{2}T)^n}{n!}\frac{e^{R|\alpha|^2}\pm 1}{2(1+e^{|\alpha|^2})},\nonumber\\
&&|\lambda^{\pm}_n\rangle=\frac{1}{\sqrt{2}}(|n\rangle_a|0\rangle_b\pm e^{in\phi}|0\rangle_a|n\rangle_b).
\end{eqnarray}
For a noon state, the quantum Fisher information is derived as $F_{q}=T^n n^2$.

In Fig. 4, we plot the phase sensitivity under a loss channel as a function of $N_{av}$, for the phase averaged states employing the modified (extended) entangled state, the entangled coherent state, and the noon state. Under the same transmission rate of the beam splitters $T$, we find that our modified (extended) entangled state can enhance the phase estimation in the small loss regime, i.e., with a large transmission $T$. 
%Note that we considered the pronounced phase terms ($\varphi=\pi/2,~\pi$) of the modified entangled state.
When the loss is further increased ($T=0.85$), the phase averaged noon state exhibits a bit better performance of the phase estimation in a large $N_{av}$. It may be attributed to the fact that the noon state is unchanged under phase averaging, which can affect robustness against optical loss. This issue may deserve further investigations in future.

\section{conclusion}
We have introduced a 4HCS, which is a superposition of four different coherent states.
We have proposed employing the entangled state produced by the 4HCS for the purpose of phase measurement, 
and shown that the 4HCS provides enhancement in phase sensitivity over the 2HCS in the regime of small average photon number $N_{av}\leq 0.7$.
To get a better performance in the regime of $N_{av}>1$, we have also proposed a modified entangled state,
$(|\alpha\rangle_a+|-\alpha\rangle_a)|0\rangle_b+|0\rangle_a(|\alpha\rangle_b+|-\alpha\rangle_b)$, which yields an improved performance
over the entangled coherent state ($|\alpha\rangle_a|0\rangle_b+|0\rangle_a|\alpha\rangle_b$) generated by the 2HCS.
Extending the superposition to $N$ different coherent states ($|C_N(\alpha)\rangle$), 
the optimal phase estimation has been further enhanced with the increasing number of component coherent states ($N$), 
which may be attributed to the fact that the number variance of one of the two input modes increases with the number of the different coherent states ($N$) under the same average photon number.
The enhancement of the optimal phase estimation has been presented even in a large average photon number.
Furthermore, we have shown that the optimal phase estimation employing the phase averaged state of the modified (extended) entangled state can perform better than that employing one of the entangled coherent state, even in the presence of loss.
For all the states considered in this paper, which belong to the class of path-symmetric state \cite{SKDL13}, the phase-estimation limit identified for each state can be achieved on a specific phase via a parity measurement scheme \cite{SKDL13}.

To identify a best possible resource for phase estimation under practical situations, we may further extend the current study to other classes of superposition states. For instance, a future work can possibly consider other types of entangled states, e.g., an arbitrary photon-number entangled state in a finite dimension \cite{LPLN12}, a broad class of non-Gaussian entangled states known to be useful for quantum information processing.

%\section{Figures and Tables}It is not necessary to place figures and tables at the back of the manuscript. Figures and tables should be sized as they are to appear in the final article. Do not include a separate list of figure captions and table titles.Figures and Tables should be labelled and referenced in the standard way using the \verb|\label{}| and \verb|\ref{}| commands.

%\subsection{Sample Figure} Figure \ref{fig:false-color} shows an example figure.
%\begin{figure}[htbp]
%\centering
%\fbox{\includegraphics[width=\linewidth]{sample}}
%\caption{False-color image, where each pixel is assigned to one of seven reference spectra.}
%\label{fig:false-color}
%\end{figure}

\section*{Acknowledgments}
S.-Y. L. thanks Changhyoup Lee, Changsuk Noh, and Kavan Modi for useful comments.
This work was supported by an NPRP grant 4-520-1-083 from Qatar National Research Fund, the National Research Foundation and Ministry of Education in Singapore.
%C.-W.L. and H.N. were supported by ...

\appendix
\section*{Appendix}
The extended entangled state is represented by
\begin{eqnarray}
|\psi_{in}\rangle=\frac{|C_N(\alpha)\rangle_a|0\rangle_b+|0\rangle_a|C_N(\alpha)\rangle_b}{\sqrt{M}},
\end{eqnarray}
where $M=2[1+\{\sum^{\infty}_{n=0}\frac{|\alpha|^{2N\cdot n}}{(N\cdot n)!}\}^{-1}]$ and $|C_N(\alpha)\rangle$ is the same as Eq. (1).
%At $N=1$, the extended entangled state is equivalent to the entangled state by the 2HCS.
After applying the phase shift operation to the input state $|\psi_{out}\rangle=(I\otimes e^{i\phi\hat{b}^{\dag}\hat{b}})|\psi_{in}\rangle$, 
we evaluate its quantum Fisher information with the quantities
\begin{eqnarray}
\langle \psi_{in}|\hat{n}^2_b|\psi_{in}\rangle&=&\frac{1}{2(1+K)}
\sum^{\infty}_{n=0}\frac{|\alpha|^{2N\cdot n}(N\cdot n)^2}{(N\cdot n)!},\nonumber\\
\langle \psi_{in}|\hat{n}_b|\psi_{in}\rangle &=&\frac{1}{2(1+K)}
\sum^{\infty}_{n=0}\frac{|\alpha|^{2N\cdot n}(N\cdot n)}{(N\cdot n)!},
\end{eqnarray}
where $K=\sum^{\infty}_{n=0}\frac{|\alpha|^{2N\cdot n}}{(N\cdot n)!}$.

We obtain a phase averaged state of the extended entangled state as
\begin{eqnarray}
\rho_m&=&\frac{1}{N^2(1+K)}\sum^{\infty}_{n=0}\frac{|\alpha|^{2n}}{n!}[N+2\sum^{N-1}_{q=1}(N-q)\cos(\frac{2\pi n}{N}q)]\nonumber\\
&&\times|noon\rangle_{ab}\langle noon|,
\end{eqnarray}
where $|noon\rangle_{ab}=(|n\rangle_a|0\rangle_b+|0\rangle_a|n\rangle_b)/\sqrt{2}$ and $K=\sum^{\infty}_{n=0}\frac{|\alpha|^{2N\cdot n}}{(N\cdot n)!}$. On applying a phase shift operation, we obtain the quantum Fisher information
\begin{eqnarray}
F_q=\frac{1}{1+K}\sum^{\infty}_{n=0}\frac{|\alpha|^{2N\cdot n}}{(N\cdot n)!}(N\cdot n)^2.
\end{eqnarray}

In the presence of loss, we obtain the output state of Eq. (22) in terms of its eigenvalues and eigenvectors as
\begin{eqnarray}
\rho_{ab}=\sum^{\infty}_{m=0}(\lambda^+_m|\lambda^+_m\rangle\langle \lambda^+_m|
+\lambda^-_m|\lambda^-_m\rangle\langle \lambda^-_m|),
\end{eqnarray}
where 
\begin{eqnarray}
&&\lambda^{+}_m=\frac{T^m}{N^2(1+K)}[C_m(N)+\frac{1}{2}\sum^{\infty}_{n=m+1}\frac{n!C_n(N)R^{n-m}}{(n-m)!m!}], \nonumber\\
&&\lambda^{-}_m=\frac{T^m}{2N^2(1+K)}\sum^{\infty}_{n=m+1}\frac{n!C_n(N)R^{n-m}}{(n-m)!m!},\nonumber\\
&&C_n(N)=\frac{|\alpha|^{2n}}{n!}[N+2\sum^{N-1}_{k=1}(N-k)\cos(\frac{2\pi n}{N}k)],\nonumber\\
&&|\lambda^{\pm}_m\rangle=\frac{1}{\sqrt{2}}(|m\rangle_a|0\rangle_b\pm e^{im\phi}|0\rangle_a|m\rangle_b),
\end{eqnarray}
with $R=1-T$ ($T$: transmission rate of the beam splitter).

% Bibliography
%\bibliography{QPE}

\begin{thebibliography}{}
  
\bibitem{MMKW96} 
  C.~Monroe, D.M.~Meekhof, B.E.~King, and D.J.~Wineland, 
  ``A ``Schr\"odinger Cat'' superposition state of an atom'', Science \textbf{272}, 1131 (1996).


\bibitem{GMHB02} 
M.~Greiner, O.~Mandel, T.W.~H\"ansch, and I.~Bloch, 
``Collapse and revival of the matter wave field of a Bose-Einstein condensate",
Nature \textbf{419}, 51 (2002). 

\bibitem{B96} 
M.~Brune et al., 
``Observing the Progressive Decoherence of the ``Meter'' in a Quantum Measurement'',
Phys. Rev. Lett. \textbf{77}, 4887 (1996).

\bibitem{PPS08} 
F.~Piazza, L.~Pezz\'e, and A.~Smerzi, 
``Macroscopic superpositions of phase states with Bose-Einstein condensates'',
Phys. Rev. A \textbf{78}, 051601(R) (2008).

\bibitem{FMH09} 
G.~Ferrini, A.~Minguzzi, and F.W.J.~Hekking, 
``Detection of coherent superpositions of phase states by full counting statistics in a Bose Josephson junction'',
Phys. Rev. A \textbf{80}, 043628 (2009).

\bibitem{Mecozzi87} 
A.~Mecozzi, and P.~Tombesi, 
``Distinguishable quantum states generated via nonlinear birefringence'',
Phys. Rev. Lett. \textbf{58}, 1055 (1987).

\bibitem{Yurke87} 
B.~Yurke and D.~Stoler, 
``Generating quantum mechanical superpositions of macroscopically distinguishable states via amplitude dispersion'',
Phys. Rev. Lett. \textbf{57}, 13 (1986), in which an input coherent state can emerge as a two-headed or a four-headed cat state at the output depending on the order of anharmonicity in the interaction Hamiltonian. 

\bibitem{Sanders92} 
B.C.~Sanders, 
``Entangled coherent states'',
Phys. Rev. A  \textbf{45}, 6811 (1992).

\bibitem{Gerry99} 
C.C.~Gerry, 
``Generation of optical macroscopic quantum superposition states via state reduction with a Mach-Zehnder interferometer containing a Kerr medium'',
Phys. Rev. A  \textbf{59}, 4095 (1999).

\bibitem{Paternostro03} 
M.~Paternostro, M.S.~Kim, and B.S.~Ham, 
``Generation of entangled coherent states via cross-phase-modulation in a double electromagnetically induced transparency regime'',
Phys. Rev. A  \textbf{67}, 023811 (2003).

\bibitem{Ourjoumtsev09} 
A.~Ourjoumtsev, F.~Ferreyrol, R.~Tualle-Brouri, and P.~Grangier, 
``Preparation of non-local superpositions of quasi-classical light states'',
Nat. Phys.  \textbf{5}, 189 (2009).

\bibitem{Brask10} 
J.B.~Brask, I.~Rigas, E.S.~Polzik, U.L.~Andersen, and A.S.~Sorensen, 
``Hybrid Long-Distance Entanglement Distribution Protocol'',
Phys. Rev. Lett. \textbf{105}, 160501 (2010).

\bibitem{Mann95} 
A.~Mann, B.C.~Sanders, W.J.~Munro, 
``Bell's inequality for an entanglement of nonorthogonal states'',
Phys. Rev. A  \textbf{51}, 989 (1995).

\bibitem{Paternostro10} 
M.~Paternostro, H.~Jeong, 
``Testing nonlocal realism with entangled coherent states'',
Phys. Rev. A  \textbf{81}, 032115 (2010).

\bibitem{McKeown11} 
G.~McKeown, M.G.A.~Paris, M.~Paternostro, 
``Testing quantum contextuality of continuous-variable states'',
Phys. Rev. A  \textbf{83}, 062105 (2011).

\bibitem{Lee13} 
C.-W.~Lee, S.-W.~Ji, H.~Nha, 
``Quantum steering for continuous-variable states'',
J. Opt. Soc. Am. B  \textbf{30}, 2483 (2013).

\bibitem{Cochrane99} 
P.T.~Cochrane, G.J.~Milburn, W.J.~Munro, 
``Macroscopically distinct quantum-superposition states as a bosonic code for amplitude damping'',
Phys. Rev. A  \textbf{59}, 2631 (1999).

\bibitem{JKL01} 
H.~Jeong, M.S.~Kim, and J.~Lee, 
``Quantum-information processing for a coherent superposition state via a mixed entangled coherent channel'',
Phys. Rev. A \textbf{64}, 052308 (2001).

\bibitem{JK02} 
H.~Jeong and M.S.~Kim, 
``Efficient quantum computation using coherent states'',
Phys. Rev. A \textbf{65}, 042305 (2002).

\bibitem{Ralph03} 
T.C.~Ralph, A.~Gilchrist, G.J.~Milburn, W.J.~Munro, and S.~Glancy, 
``Quantum computation with optical coherent states'',
Phys. Rev. A  \textbf{68}, 042319 (2003).

\bibitem{Lund08} 
A.P.~Lund, T.C.~Ralph, H.L.~Haselgrove, 
``Fault-Tolerant Linear Optical Quantum Computing with Small-Amplitude Coherent States'',
Phys. Rev. Lett. \textbf{100}, 030503 (2008).

\bibitem{Kim10} 
J.~Kim, J.~Lee, S.-W.~Ji, H.~Nha, P.M.~Anisimov, and J.P.~Dowling, 
``Coherent-state optical qudit cluster state generation and teleportation via homodyne detection'',
Opt. Commun. \textbf{337}, 79 (2015).

\bibitem{Lund11} 
C.R.~Myers and T.C.~Ralph, 
``Coherent state topological cluster state production'',
New J. Phys.  \textbf{13}, 115015 (2011).

\bibitem{Neergaard13} 
J.S.~Neergaard-Nielsen, Y.~Eto, C.-W.~Lee, H.~Jeong, and M.~Sasaki, 
``Quantum tele-amplification with a continuous variable superposition state'',
Nat. Photon.  \textbf{7}, 439 (2013).

\bibitem{Sangouard10} 
N.~Sangouard et al., 
``Quantum repeaters with entangled coherent states'',
J. Opt. Soc. Am. B  \textbf{27}, A137-A145 (2010).

\bibitem{Ralph09} 
T.C.~Ralph and G.J.~Pryde, 
``Optical quantum computation'',
Prog. Opt.  \textbf{54}, 209 (2009).

\bibitem{Munro02} 
W.J.~Munro, K.~Nemoto, G.J.~Milburn, and S.L.~Braunstein, 
``Weak-force detection with superposed coherent states'',
Phys. Rev. A  \textbf{66}, 023819 (2002).

\bibitem{Joo11} 
J.~Joo, W.J.~Munro, and T.P.~Spiller, 
``Quantum Metrology with Entangled Coherent States'',
Phys. Rev. Lett. \textbf{107}, 083601 (2011).

\bibitem{Y13} 
M.~Yukawa et al., 
``Generating superposition of up-to three photons for continuous variable quantum information processing'',
Opt. Express \textbf{21}, 5529 (2013).

\bibitem{R10} 
J.M.~Raimond et al., 
``Phase Space Tweezers for Tailoring Cavity Fields by Quantum Zeno Dynamics'',
Phys. Rev. Lett. \textbf{105}, 213601 (2010).

\bibitem{V13} 
B.~Vlastakis et al., 
``Deterministically Encoding Quantum Information Using 100-Photon Schr\"odinger Cat States'',
Science \textbf{342}, 607 (2013).

\bibitem{L13} 
Z.~Leghtas et al., 
``Hardware-Efficient Autonomous Quantum Memory Protection'',
Phys. Rev. Lett. \textbf{111}, 120501 (2013).

\bibitem{R12} 
J.M.~Raimond et al., 
``Quantum Zeno dynamics of a field in a cavity'',
Phys. Rev. A \textbf{86}, 032120 (2012).


\bibitem{LJL13} 
S.-Y.~Lee, S.-W.~Ji, and C.-W.~Lee, 
``Increasing and decreasing entanglement characteristics for continuous variables by a local photon subtraction'',
Phys. Rev. A \textbf{87}, 052321 (2013).

\bibitem{C81} 
C.M.~Caves, 
``Quantum-mechanical noise in an interferometer'',
Phys. Rev. D \textbf{23}, 1693 (1981).

\bibitem{D08} 
J.P.~Dowling, 
``Quantum optical metrology - The lowdown on high-N00N states'',
Contemp. Phys. \textbf{49}, 125 (2008).


\bibitem{ZPK10} 
M.~Zwierz, C.A.~P\'erez-Delgado, and P.~Kok, 
``General Optimality of the Heisenberg Limit for Quantum Metrology'',
Phys. Rev. Lett. \textbf{105}, 180402 (2010). 

\bibitem{J12} 
J.~Joo et al., 
``Quantum metrology for nonlinear phase shifts with entangled coherent states'',
Phys. Rev. A \textbf{86}, 043828 (2012).

\bibitem{JD12} 
M.~Jarzyna and R.~Demkowicz-Dobrza\'nski, 
``Quantum interferometry with and without an external phase reference'',
Phys. Rev. A \textbf{85}, 011801 (R) (2012).

\bibitem{Bowen}  
M.A.~Taylor and W.P.~Bowen, 
``Quantum metrology and its application in biology'',
arXiv:1409.0950

%\bibitem{PHS14} 
%L.~Pezz\`e, P.~Hyllus, and A.~Smerzi, 
%``Phase sensitivity bounds for two-mode interferometers'',
%arXiv:1408.6971

\bibitem{KSD11} 
S.~Knysh, V.N.~Smelyanskiy, and G.A.~Durkin, 
``Scaling laws for precision in quantum interferometry and the bifurcation landscape of the optimal state'',
Phys. Rev. A \textbf{83}, 021804 (R) (2011).

\bibitem{ZLYJ13} 
Y.M.~Zhang, X.W.~Li, W.~Yang, and G.R.~Jin, 
``Quantum Fisher information of entangled coherent states in the presence of photon loss'',
Phys. Rev. A \textbf{88}, 043832 (2013).

\bibitem{LJW13} 
J.~Liu, X.~Jing, and X.~Wang, 
``Phase-matching condition for enhancement of phase sensitivity in quantum metrology'',
Phys. Rev. A \textbf{88}, 042316 (2013).

\bibitem{SKDL13}
K.P.~Seshadreesan, S.~Kim, J.P.~Dowling, and H.~Lee,
``Phase estimation at the quantum Cram\'er-Rao bound via parity detection",
Phys. Rev. A \textbf{87}, 043833 (2013).

\bibitem{LPLN12} 
S.-Y.~Lee, J.~Park, H.-W.~Lee, and H.~Nha, 
``Generating arbitrary photon-number entangled states for continuous-variable quantum informatics'',
Opt. Express \textbf{20}, 14221 (2012).
  
\end{thebibliography}

%Manual citation list

\end{document}